\documentclass[preprint,3p,times,onecolumn,hyphens]{elsarticle}
\biboptions{sort&compress}
\usepackage{amsmath}
\usepackage{amssymb}
\usepackage{xfrac}
\usepackage{siunitx}

\usepackage{booktabs}
\usepackage{multirow}
\usepackage{multicol}
\usepackage{tabularray}
\UseTblrLibrary{booktabs, amsmath}

\usepackage{float}       
\usepackage{placeins}    
\usepackage{afterpage}   
\usepackage{caption}
\usepackage{subcaption}

\usepackage{hyperref}
\hypersetup{colorlinks=true,hidelinks}
\usepackage{cleveref}    
\usepackage{url}         

\usepackage{framed}
\usepackage[prefix,nostdsubgroups]{nomencl}
\makenomenclature
\renewcommand*\nompreamble{\begin{multicols}{2}}
\renewcommand*\nompostamble{\end{multicols}}

\usepackage{xcolor}

\journal{Computer Methods in Applied Mechanics and Engineering}

\begin{document}

\begin{abstract}

Wire-arc directed energy deposition (DED) has emerged as a promising additive manufacturing (AM) technology for large-scale applications in structural engineering. However, the complex thermal dynamics inherent to the process present challenges in ensuring structural integrity and mechanical properties of fabricated thick wall and plates. While finite element method (FEM) simulations have been conventionally employed to predict thermal history during deposition, their high computational demand remains prohibitively high for actual large-scale applications. Given the necessity of multiple repetitive simulations for heat management and the determination of optimal printing strategy, FEM simulation quickly becomes entirely infeasible and unfit. Instead, advancements have been made in using trained neural networks as surrogate models for rapid prediction. However, traditional data-driven approaches necessitate large amounts of relevant and verifiable external data, either from simulation, experimental, or analytical solutions, during the training and validation of the neural network. Regarding large-scale wire-arc DED, none of these data sources are readily available in quantities sufficient for an accurate surrogate. The introduction of physics-informed neural networks (PINNs) has opened up an alternative simulation strategy by leveraging the existing physical knowledge of the phenomena with advanced machine learning methods. However, the practical application of PINNs for wire-arc DED has been rarely explored, particularly within the context of structural engineering, where large-scale metal AM is demanded. This study investigates the necessary steps for upscaling PINN with a focus on advanced and effective sampling of collocation points — a critical factor controlling both the training time and the performance of the model. The results affirm the potential of PINNs to outperform FEM, with marked reduction in computational time and effort up to 98.6\%, while maintaining the desired accuracy and offering "super-resolution". Further discussion provides an outlook on the future steps for improving the PINNs for wire-arc DED simulations.

\end{abstract}

\begin{frontmatter}
    
    \title{Physics-informed machine learning surrogate for scalable simulation of thermal histories during wire-arc directed energy deposition}
    \author[luh]{Michael Ryan}
    \author[luh]{Mohammad Hassan Baqershahi}
    \author[luh]{Hessamoddin Moshayedi\corref{cor1}}\ead{moshayedi@stahl.uni-hannover.de}
    \author[luh]{Elyas Ghafoori}

    \cortext[cor1]{Corresponding author}
    \affiliation[luh]{organization={Institute for Steel Construction},
                 addressline={Leibniz University Hannover},
                 city={Hannover},
                 postcode={30167},
                country={Germany}
    }
\end{frontmatter}

\begin{table*}[!t]   
    \begin{framed}
        \nomenclature{PINN}{Physics-Informed Neural Network}
\nomenclature{FEM}{Finite Element Method}
\nomenclature{AM}{Additive Manufacturing}
\nomenclature{DED}{Directed Energy Desposition}
\nomenclature{CFD}{Computational Fluid Dynamics}
\nomenclature{PDE}{Partial Differential Equation}
\nomenclature{MLP}{Multilayer Perceptron}
\nomenclature{AD}{Automatic Differentiation}
\nomenclature{BC}{Boundary Condition}
\nomenclature{$\rho$}{Density}
\nomenclature{$C_{p}$}{Specific Heat}
\nomenclature{$k$}{Thermal Conductivity}
\nomenclature{$Q_{goldak}$}{Volumetric heat flux}
\nomenclature{$P$}{Power}
\nomenclature{$\eta$}{Weld efficiency}
\nomenclature{$a$}{Length of melt pool}
\nomenclature{$b$}{Width of melt pool}
\nomenclature{$c$}{Depth of melt pool}
\nomenclature{$f_{f}$}{Front distribution factor of heat flux}
\nomenclature{$f_{r}$}{Rear distribution factor of heat flux}
\nomenclature{$R_{PDE}$}{Heat conduction residual}
\nomenclature{$R^{(-z)}_{BC}$}{Dirichlet boundary condition residual}
\nomenclature{$R_{BC}$}{Robin boundary condition residual}
\nomenclature{$T_{0}$}{Initial/Ambient material temperature}
\nomenclature{$\sigma$}{Stefan-Boltzmann constant}
\nomenclature{$q_{conv}$}{Convective heat transfer}
\nomenclature{$q_{rad}$}{Radiative heat transfer}
\nomenclature{$\varepsilon$}{Material emissivity}
\nomenclature{$\mathcal{L}_{BC}$}{BC Loss}
\nomenclature{$\mathcal{L}_{PDE}$}{PDE Loss}
\nomenclature{$\mathcal{L}$}{Total Loss}
\nomenclature{$w_{BC}$}{BC Loss weight}
\nomenclature{$w_{PDE}$}{PDE Loss weight}
\nomenclature{GELU}{Gaussian Error Linear Unit}
\nomenclature{ReLU}{Rectified Linear Unit}
\nomenclature{$\hat{u}$}{Neural Network output}
\nomenclature{L-BFGS}{Limited Memory BFGS}
\nomenclature{RWF}{Random Weight Factorization}
        \printnomenclature
    \end{framed}
\end{table*}

\section{Introduction}\label{sec:introduction}

\begin{figure}[H]
    \centering
    \includegraphics[width=0.7\linewidth]{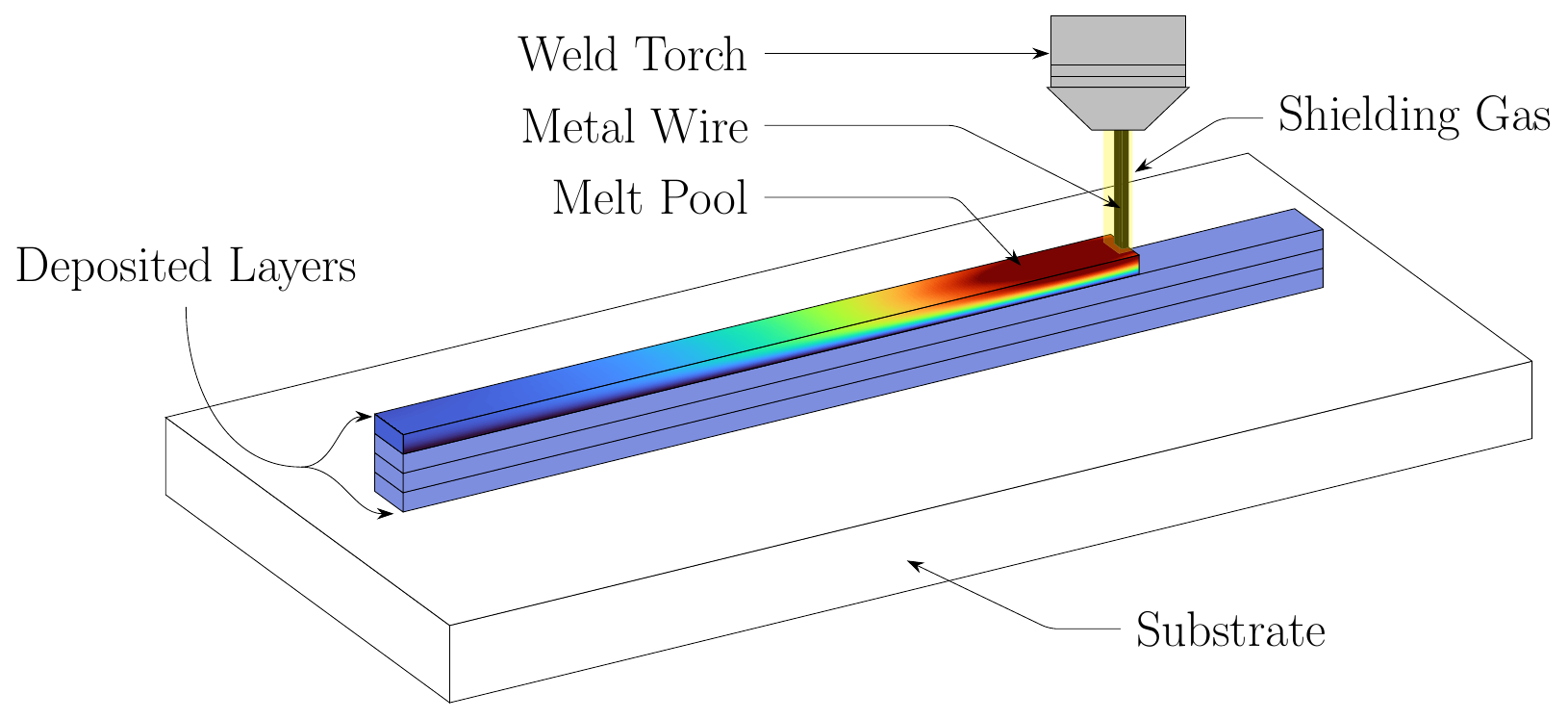}
    \caption{Schematic representation of wire-arc DED}
    \label{fig:waam_schematic}
\end{figure}

\begin{figure}[H]
    \centering
    \includegraphics[width=0.6\linewidth]{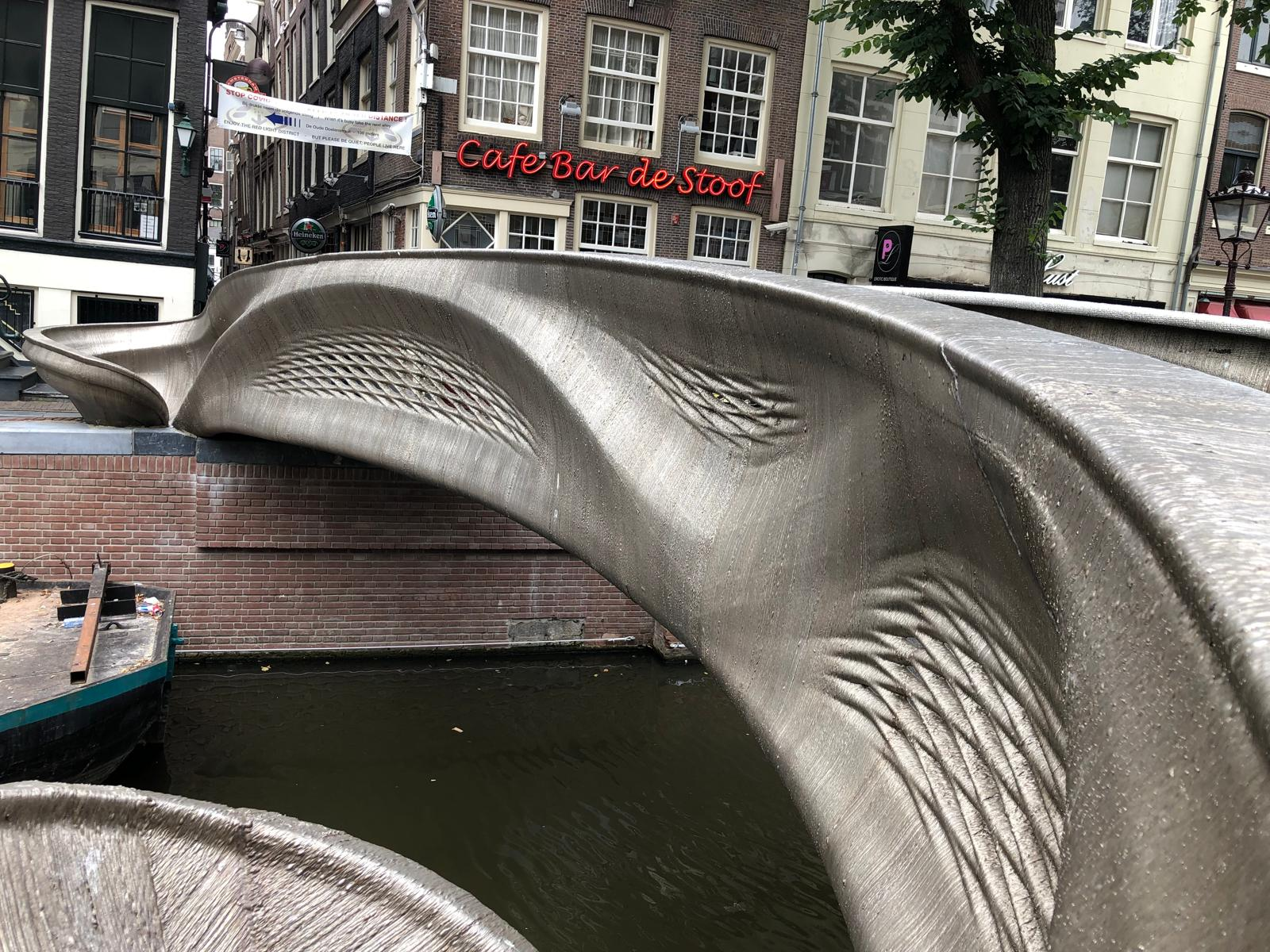}
    \caption{MX3D: Stainless Steel Additive-Manufactured Bridge - made through wire-arc DED - Amsterdam, NL}
    \label{fig:mx3d}
\end{figure}

Wire-arc directed energy deposition (DED) is an additive manufacturing (AM) technique based on conventional welding technology, utilizing a robotic system for the layer-by-layer deposition of complex structures (\Cref{fig:waam_schematic}). It has shown a great potential for fabricating medium- to large-scale components, making it the most promising metal AM technique for structural applications~\cite{buchanan2019metal}, owing to its lower costs and higher deposition rates compared with other technologies such as powder bed fusion~\cite{williams2016wire+}. By enabling the fabrication of innovative, efficient, and complex structures that were previously unachievable, it can enhance sustainability in construction~\cite{baqershahi2024design, baqershahi2025topology}. It also opens up the doors to advanced materials, such as shape memory alloys, for the design of novel structures~\cite{jafarabadi20234d, felice2023wire, lopes2024unveiling}.
Wire-arc DED has already been used for strengthening of conventional profiles~\cite{yang2025strengthening,meng2025hybrid,dahaghin2024wire,ghafoori2023fatigue} and for manufacturing of structural components such as columns~\cite{laghi2020computational}, beams~\cite{ye2021end}, and nodes~\cite{wang2021joints,snijder2020glass}, as well as entire structures~\cite{gardner2020testing,HUANG2023103696,HUANG2023107705}. What distinguishes the application of wire-arc DED in structural engineering from other fields is the scale of components that can be up to meters, as shown in~\Cref{fig:mx3d}.
While opening up space for unprecedented design freedom, wire-arc DED also raises several open questions regarding mechanical properties, the structural reliability, and certification, among others~\cite{gardner2023metal}. An intrinsic complexity of the wire-arc DED technique lies in its unique temperature history, characterized by high thermal and cooling rates. These factors directly influence the common defects, microstructure, and mechanical properties of the component, potentially resulting in complex residual stress distributions and distortions~\cite{rodrigues2019wire, dahaghin2024wire}, which can result in inferior quality and even total failure~\cite{sames2016metallurgy}. These effects are pronounced on larger scales, making it necessary to identify the optimal process parameters and effectively manage temperature evolution to mitigate defects and ensure the mechanical properties required for specific applications. However, experimental trial and error is a cost barrier preventing the widespread adoption of this technology in construction. As a result, numerical simulations become a viable tool for moving towards the goal of first-time-right, high-quality production of large-scale components.

Two primary physics-based computational methods used to simulate the metal AM processes are finite element method (FEM) and computational fluid dynamics (CFD). CFD simulations enable a more accurate study of the melt pool by considering thermal-fluid interactions~\cite{mukherjee2018heat}; however, they are limited to thermal analysis, and due to the associated high computational cost, they are often only utilized at the mesoscale~\cite{liaoHybridThermalModeling2023}. FEM models, capable of both thermal and mechanical analysis, consider heat transfer in solid state and are widely used for macroscopic simulations to predict, for instance, distortions and residual stresses~\cite{lu2019residual, dahaghin2024wire}. However, the computational cost for thermal simulations using FEM still increases rapidly with the scale of the problem~\cite{yang2021physics}, not because of the complexity of the physical phenomena, but rather due to the localized nature of the problem with high temperature gradients~\cite{hosseini2023single}. This renders the thermal simulation of medium- to large-scale components relevant for structural engineering practically impossible, necessitating alternative simulation strategies.

Along with the rapid development of machine learning (ML) algorithms, a new paradigm known as scientific ML has emerged, distinguishing itself from purely data-driven approaches. They attempt to integrate domain-specific knowledge and theoretical understanding, rather than relying solely on large datasets, to leverage the strengths of both approaches, thereby enhancing accuracy and interpretability~\cite{karniadakis2021physics}.

A particular representation of this philosophy is the family of physics-informed neural networks (PINNs), which leverages a deep learning framework to solve partial differential equations (PDEs). This is achieved by defining the loss function based on the residual of the governing PDE, where automatic differentiation (AD) is used to calculate the derivative operators of the PDE without the discretization error typical of numerical differentiation.

One of the problems that has received attention is the heat transfer problem, which is also relevant for the prediction of thermal history during the wire-arc DED process. Zhu et al.~\cite{zhu2021machine} developed a PINN to predict temperature, velocity, and pressure fields. A hard-way implementation of Dirichlet boundary conditions was carried out to ensure their fulfillment and accelerate the learning process. Manufacturing parameters, such as material properties, laser power, and scanning speed, were assumed to be constant. Validations were also performed by comparing the results with analytical solutions for 1D problems and validated FE models. Xie et al.~\cite{xie20223d} developed a PINN model for predicting temperature fields for directed energy deposition. They included laser power and scanning speed, in addition to spatial-temporal coordinates, as the inputs. The extension to multilayer deposition was made for three specific cases, considering different scanning strategies and preheating conditions. The results were verified against FE models and measured data. Hosseini et al.~\cite{hosseini2023single} developed a parametric PINN for a single-track thermal simulation of the laser power bed fusion (LPBF) process, incorporating several additional inputs, including heat input, scanning speed, and the material’s thermal properties. Liao et al.~\cite{liaoHybridThermalModeling2023} developed a PINN for temperature prediction with only spatiotemporal coordinates and applied it to an inverse problem to identify unknown materials and process parameters. They reported that adding auxiliary labeled data for training could accelerate the training process, even if the data is noisy and of low quality. Uhrich et al.~\cite{uhrich2024physics} developed a PINN that considered phase transformations between liquid and solids. They also utilized transfer learning as a computationally efficient approach to extend the capability of PINN to a broader range of heat inputs, as well as for multilayer deposition.

Despite the extensive research on PINNs, the primary focus has been on proving of concept, assessing its feasibility, and comparing the prediction accuracy with that of conventional methods. However, a noticeable gap exists in the literature that is required to make this numerical approach suitable for real-world, large-scale applications. This study aims to lay the groundwork for upscaling simulations by utilizing more advanced strategies in training PINN and quantitatively comparing the computational efficiency of PINN in comparison with that of FE for thermal simulations in large-scale components relevant to structural engineering.

\section{Methodology}\label{sec:methodology}
In contrast to data-driven approaches, the basic principle of physics-informed neural networks is to utilize our acquired knowledge of physical phenomena to constrain the network to physical laws. One can impose soft constraints on the network by simply including residuals of relevant partial differential equations, initial conditions, and boundary conditions into the loss function. By calculating the residual of the physical equations using the current model prediction, the deviation from expected physical constraints can be used to update weights and refine the model. Without external data, the model is driven entirely by physical constraints.

\subsection{Governing Equations}\label{subsec:governing_equations}
The heat conduction residual equation can be defined as: 
\begin{equation}
    R_{PDE} = \rho C_{p}\dfrac{\partial T}{\partial t} - k\nabla^{2}T - Q_{goldak}(\mathbf{x}, t)
\end{equation}
where $C_{p}$, $\rho$, and $k$ are the specific heat, density, and thermal conductivity of the material, respectively, $T$ is the temperature at ($\mathbf{x}, t$), and $Q_{goldak}$ is the volumetric heat flux.

\begin{figure}[h]
    \centering
    \includegraphics[trim={0.35cm 0 0 0}, clip, width=0.5\linewidth]{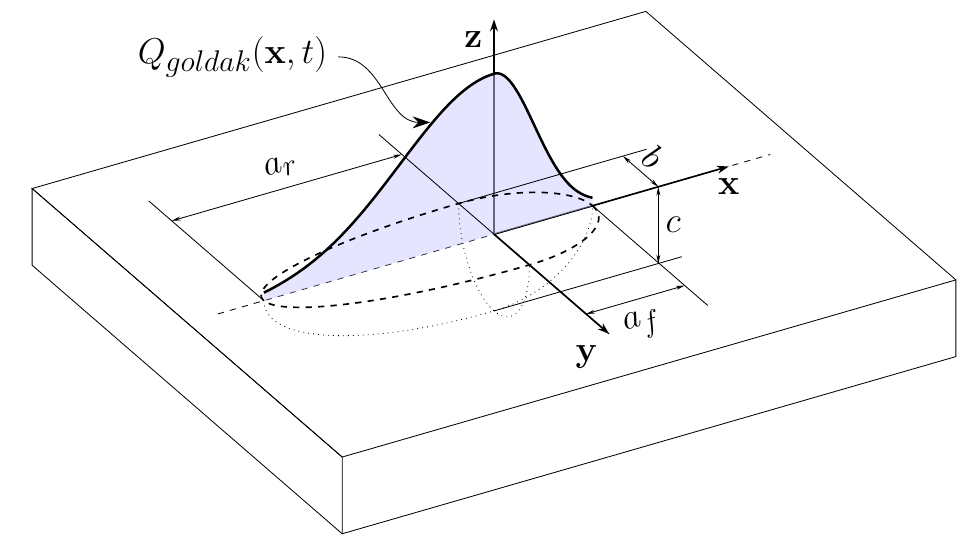}
    \caption{Goldak's double-ellipsoidal heat distribution}
    \label{fig:goldak_heat_source}
\end{figure}

As it has proven most representative for wire-arc DED~\cite{SAMPAIO2023100121}, the Goldak heat source is defined piecewise as~\cite{goldakNewFiniteElement1984}:  
\begin{equation}\label{eq:goldak_heat_source}
    Q_{goldak} =
        \begin{+cases}
        \frac{6\sqrt{3} f_f P \eta}{a_f \cdot b \cdot c \cdot \pi \sqrt{\pi}} \exp\left(-3\left(\frac{x^2}{a_f^2} + \frac{y^2}{b^2} + \frac{z^2}{c^2}\right)\right), & x \leq 0  \\
        \\
         \frac{6\sqrt{3} f_r P \eta}{a_r \cdot b \cdot c \cdot \pi \sqrt{\pi}} \exp\left(-3\left(\frac{x^2}{a_r^2} + \frac{y^2}{b^2} + \frac{z^2}{c^2}\right)\right), & x > 0 
        \end{+cases}
\end{equation}
where $f_{f}$ and $f_{r}$ are the front and rear distribution factors, respectively, $P$ is the power of the source, and $\eta$ is the arc efficiency. The parameters $a$, $b$, and $c$ represent the melt pool dimensions: length, width, and depth, respectively. The subscripts $f$ and $r$ denote the front and rear half-lengths of the melt pool along the direction of travel, as illustrated in \Cref{fig:goldak_heat_source}. To simulate movement, the equation is modified to include explicit time dependence. For our purposes, the heat source travels at a uniform velocity $v$ along the $x$-axis; therefore, the $x$ component of \Cref{eq:goldak_heat_source} is replaced by $x - (x_{0} + v\cdot t)$.

The heat flux boundary condition residual consists of two parts: convection and radiation, applied to each surface as necessary. 
\begin{equation}
    R_{BC} = -k\dfrac{\partial T}{\partial  \vec{n}} - (q_{conv} + q_{rad})
\end{equation}
The convective and radiative heat transfer are calculated, respectively, as:
\begin{equation}\label{eq:convective_heat_flux}
    q_{conv} = h(T - T_{0})
\end{equation}
\begin{equation}\label{eq:radiation_heat_flux}
    q_{rad} = \sigma\varepsilon(T^{4} - T_{0}^{4})
\end{equation}
where $h$ is the convection coefficient, $\sigma$ is the Stefan-Boltzmann constant, $\varepsilon$ is the material emissivity, and $T_{0}$ is the ambient temperature. The Robin boundary condition detailed above is applied to all surfaces except the bottom surface ($-z$), in which a simple Dirichlet boundary condition is imposed as
\begin{equation}
    R_{BC}^{(-z)} = T - T_{0}.
\end{equation}

\subsection{Loss}
The model is trained by minimizing the normalized weighted total loss $\mathcal{L}$, calculated as 
\begin{equation}
    \mathcal{L}_{BC} = \frac{1}{N_{BC}}\sum_{i=1}^{N_{BC}} \left| R_{BC}(x_{BC}^{i}, t_{BC}^{i}) \right|^{2}
\end{equation}

\begin{equation}
    \mathcal{L}_{PDE} = \frac{1}{N_{PDE}}\sum_{i=1}^{N_{PDE}} \left| R_{PDE}(x_{PDE}^{i}, t_{PDE}^{i}) \right|^{2}
\end{equation}

\begin{equation}
    \mathcal{L} = \dfrac{w_{BC}\mathcal{L}_{BC} + w_{PDE}\mathcal{L}_{PDE}}{w_{BC} + w_{PDE}}
\end{equation}
with $w_{BC}$ and $w_{PDE}$, denoting the boundary and PDE weights, respectively. The weights are updated using a self-adaptive loss weighting scheme based on the magnitude of back-propagated gradients, following methodology by Wang et al.~\cite{wangExpertsGuideTraining2023},
\begin{equation}
    \hat{w}_{x} = \dfrac{\sum_{i=1}^{N}\left\lvert\left\lvert \nabla\mathcal{L}_i\right\rvert\right\rvert}{\left\lvert\left\lvert \nabla\mathcal{L}_x\right\rvert\right\rvert}
\end{equation}
with the loss weights then updated, as
\begin{equation}
    w_{x} = \alpha w_{x, old} + (1- \alpha)\hat{w}_{x}
\end{equation}
where $x$ represents the respective weight (${BC}$, ${PDE}$). In accordance with recommendations by Wang et al., the balance $\alpha$ was chosen as $0.9$, and the frequency of weight updating was set at 1000 epochs.

\subsection{Neural Network Setup}\label{subsec:nn_setup}
\begin{figure}[hbtp]   %
    \centering
    \includegraphics[width=0.9\linewidth]{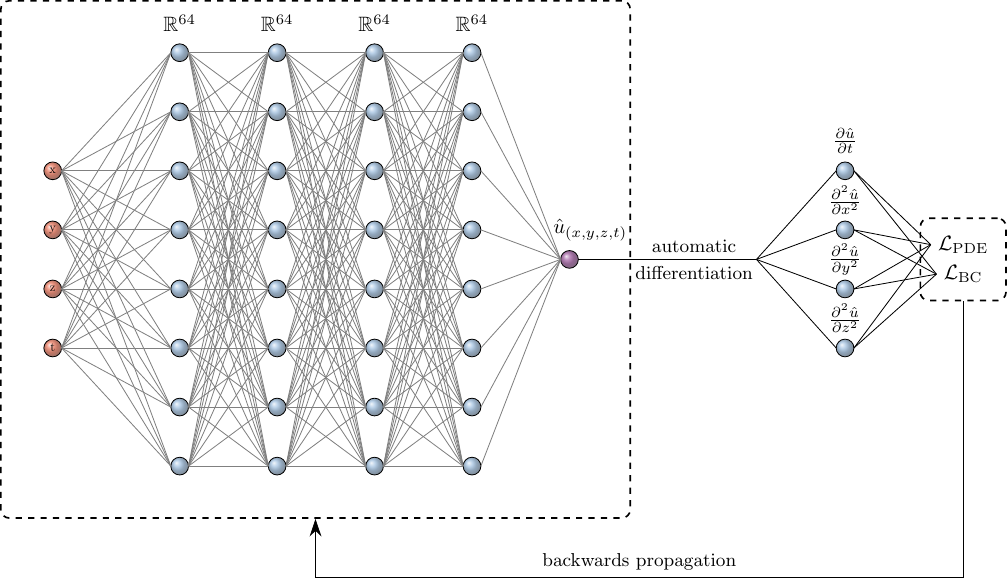}
    \caption{Network Topology- 4 Dense Layers with 64 Neurons}
    \label{fig:network_topology}
\end{figure}

The network is a fully connected, feed‐forward multilayer perceptron (MLP), as detailed in \Cref{fig:network_topology}, that maps four inputs $(x,y,z,t)$ to a single scalar output $\hat{u}(x,y,z,t)$. It has four dense hidden layers with 64 neurons. Each hidden layer uses the Gaussian Error Linear Unit (GELU), which empirical testing has shown to provide high accuracy at minimal computational cost~\cite{hendrycksGaussianErrorLinear2023}. Additionally, GELU is $C^{2}$-smooth, which is an important property when computing second‐order derivatives for the PDE residual. Since GELU activations are not zero-centered, weights are initialized using the Kaiming method with a modified gain factor of 1.48, explicitly adapted for GELU from the original ReLU-based approach~\cite{heDelvingDeepRectifiers2015}. The spatial-temporal domain inputs are sampled via a Sobol’ sequence and linearly normalized to the range $[-1, 1]$. Finally, the raw network output is passed through a Softplus activation to ensure positivity, then scaled by the expected temperature range and offset by the ambient temperature. The first- and second-order derivatives of $\hat{u}$ with respect to $x$, $y$, $z$, $t$ are computed using automatic differentiation \cite{baydinAutomaticDifferentiationAlgorithms2014}. These are used to calculate the physical losses: $\mathcal{L}_{BC}$ and $\mathcal{L}_{PDE}$. The weights are then updated using the hybrid optimizer schedule, with a static learning rate, and training proceeds until the combined loss converges or the maximum number of iterations is reached. The hybrid optimizer schedule consists of two stages: initial training with the Adam optimizer~\cite{kingmaAdamMethodStochastic2017}, followed by refinement using a limited number of L-BFGS iterations~\cite{liu1989limited}.
\subsubsection{Initial Conditions}
The initial conditions are enforced using hard constraints following methodology detailed in Roy et al.~\cite{royExactEnforcementTemporal2024}. Specifically, the output of the neural network $\hat{u}$ is further transformed in such a way that satisfies the initial conditions:
\begin{equation}
    \hat{u}_{new} = g_{0}(x,y,z) + t(\hat{u} - g_{0}(x,y,z))
\end{equation}
where $g_{0}(x,y,z)$ is the given initial condition. In our case, the initial condition is a uniform temperature field at the ambient temperature. To avoid artificially inflating predicted temperatures, the initial temperature field $g_{0}(x,y,z)$ is subtracted from the network output. By imposing strict initial conditions, we simultaneously ensure correct predictions at the initial time step and reduce the number of loss terms from three to two. This reduction significantly decreases the complexity of the optimization and allows for better results in fewer epochs for both $\mathcal{L}_{BC}$ and $\mathcal{L}_{PDE}$.

\subsection{Finite Element Modelling}\label{sec:fem}
\begin{figure}[H]
    \centering
    \includegraphics[width=0.6\linewidth]{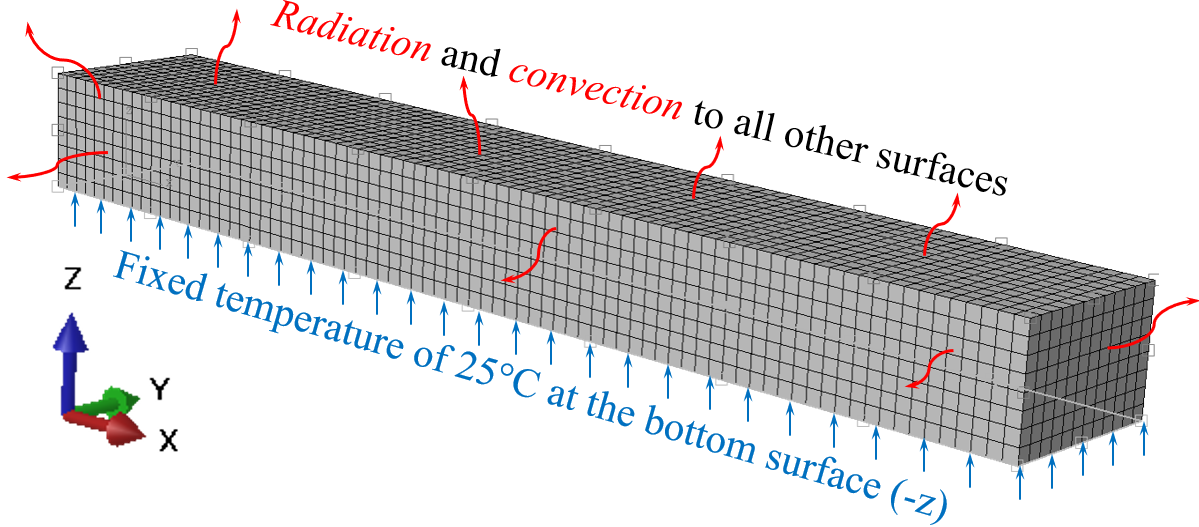}
    \caption{FEM model and mesh configuration with a mesh size of 0.5 mm}
    \label{fig:fem_model}
\end{figure}

The FEM numerical simulation was carried out using Abaqus software 2023~\cite{0b112d0e5eba4b7f9768cfe1d818872e}, purely as a benchmark analysis independent of the training process. An implicit time integration method with Newton-Raphson iteration was employed to ensure convergence in the FE analysis. The simulation was ran on four cores of the CPU Intel(R) Core(TM) Ultra 5 125U, 3.60 GHz CPU on a personal laptop with 16 GB RAM. The numerical model has already been used and validated against experiments in the authors' previous study~\cite{dahaghin2024wire}. Wire-arc DED material was deposited as a single layer in the form of a block measuring 40~mm × 6~mm × 4~mm. The FEM model, applied boundary conditions, and mesh configuration for the mesh size of 0.5 mm are shown in \Cref{fig:fem_model}.

A transient heat transfer simulation was conducted using DC3D8 elements to model the temperature evolution during deposition. Two models were prepared and analyzed, featuring mesh sizes of 0.5~mm (\Cref{fig:fem_model}) and 0.1~mm, comprising a total of 7,680 and 960,000 elements, respectively. Heat input was applied using the Goldak double ellipsoidal heat source~\cite{ueda1975new} through the DFLUX user subroutine. Goldak heat source parameters illustrated in \Cref{tab:laser_properties} were adopted from the literature by Liang et al.~\cite{liang2019modified}. Convection and radiation boundary conditions were applied using material parameters, as per \Cref{tab:material_properties}, following recommendations in Ding et al.~\cite{ding2012thermo}.
\section{Results and Discussion}\label{sec:results}

 \subsection{Problem Description}\label{subsec:problem_description}
\begin{figure}
    \centering
    \includegraphics[width=0.6\linewidth]{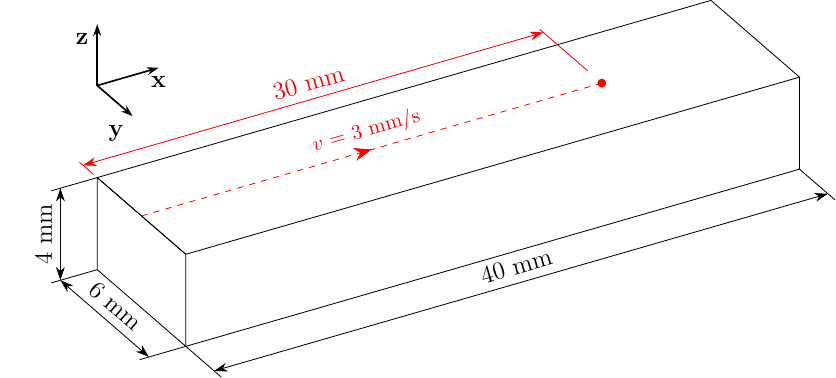}
    \caption{Schematic representation of the simplified test scenario}
    \label{fig:case1_schematic}
\end{figure}

\begin{figure}[h]
    \centering
    \includegraphics[width=0.6\linewidth, trim={5.5cm 4.5cm 3cm 5.5cm},clip]{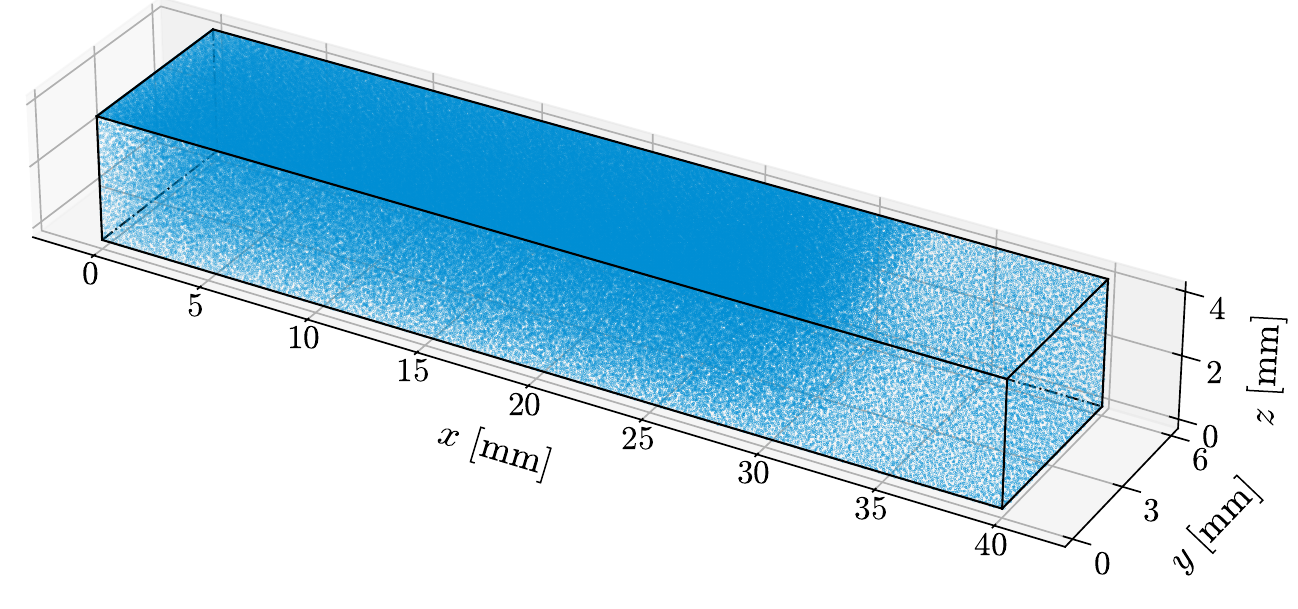}
    \caption{Sobol' sequence sampled collocation points with collapsed time dimension}
    \label{fig:case1_sampling}
\end{figure}

To evaluate the described methodology, a simplified test scenario was developed. The object domain is a bare plate with dimensions 40~mm $\times$ 6~mm $\times$ 4~mm. A Goldak heat source travels along the $x$-axis at a uniform velocity $v$ for 3 seconds, as shown in \Cref{fig:case1_schematic}. For training the physically-informed neural network, time was discretized at 5~ms intervals, and collocation points were sampled using Sobol' sequences (\Cref{subsubsec:qmc}). The points were sampled in batches and labeled according to their spatial and temporal locations in the categories of \textit{initial}, \textit{domain}, and \textit{boundary}. Additional points were sampled beneath the heat source, following its temporal location. There were 185,669 boundary points, 112,635 domain points, and 3,509 initial points for a total of 301,813 collocation points, as shown in \Cref{fig:case1_sampling}. The number of Sobol' sequence points was roughly chosen as a fraction of the estimated points required by a uniform grid of the same domain. Given higher expected gradients in the top layers, a power-law z-warp was applied on all domain points to shift them towards the top plate surface. The PINN model was implemented in PyTorch~\cite{paszkePyTorchImperativeStyle2019} and trained entirely using governing-equation-based loss functions without any external ground-truth data. The training was performed on a personal computer equipped with an NVIDIA GeForce RTX 3070Ti (8 GB VRAM) and 32 GB of system RAM. 

{\renewcommand{\arraystretch}{1.1}
\begin{table}[htbp]
    \caption{Deposition Properties}
    \vspace{-1.5mm}
    \centering
    \begin{tabular}{c l c r c}
        \toprule
                                            & \textit{Name}       & \textit{Symbol}    & \textit{Value}              & \textit{Units} \\
        \midrule
        \multirow{2}{*}{Initial}             & $x$-position       & $x_{0}$   & 0.0                 & \unit{\milli\metre} \\
                                             & $y$-position        & $y_{0}$   & 3.0                 & \unit{\milli\metre} \\
        \midrule
        \multirow{3}{*}{Process} & Velocity   & $v$       & 10.0                & \unit{\milli\metre\per\second} \\
        & Power                               & $P$       & $2.45\times10^{12}$ & \unit{\gram.\milli\metre\per\second^2} \\
        
                                              & Efficiency   & $\eta$    & 0.9                 & - \\
        \midrule
        \multirow{6}{*}{Goldak}               &\multirow{2}{*}{Length}  & $a_{f}$   & 2.57                & \unit{\milli\metre} \\
                                             &                         &  $a_{r}$   & 6.0                 & \unit{\milli\metre} \\
                                             & Width    & $b$       & 6.0                 & \unit{\milli\metre} \\
                                             & Depth     & $c$       & 4.0                 & \unit{\milli\metre} \\
        & \multirow{2}{*}{Dist. Factors} & $f_{f}$   & 0.6                 & - \\
                                             & & $f_{r}$   & 1.4                 & - \\
        \bottomrule
    \end{tabular}
    \label{tab:laser_properties}
\end{table}
}

{\renewcommand{\arraystretch}{1.5}
\begin{table}[htbp]
    \caption{Material Properties}
    \vspace{-1.5mm}    
    \centering
    \sisetup{exponent-mode = scientific, table-alignment-mode= marker}
    \begin{tabular}{l c r c}
        \toprule
        \textit{Name}             & \textit{Symbol} & \textit{Value}    & \textit{Units} \\
        \midrule
        Specific Heat            & $C_{p}$       & \num{6.20e8}   & \unit{\milli\metre^2\per\second^2.\degreeCelsius}\\
        Density                  & $\rho$        & \num{7.85e-3}  & \unit{\gram.\per\milli\metre^3} \\
        Thermal Conductivity     & $k$           & \num{4.5e7}    & \unit{\gram.\milli\metre\per\second^3.\degreeCelsius} \\
        Radiation Coefficient    & $\varepsilon$ & \num{0.2}                  & - \\
        Convection Coefficient   & $h$           & \num{2e4}      & \unit{\gram.\per\second^3.\degreeCelsius} \\
        Ambient Temperature      & $T_{0}$       & \num{25}                  & \unit{\degreeCelsius} \\
        \bottomrule
    \end{tabular}
    \label{tab:material_properties}
\end{table}
}

\subsection{Towards Large-Scale Domains}\label{subsec:speed_up} 
As the overall goal for this and subsequent publications is the simulation of large-scale processes, computational efficiency and scalability are paramount. For PINNs, this requires minimizing the number of training iterations, reducing the per-iteration computational cost, and, critically, addressing the exponential growth in sampling complexity with dimensionality $m^d$.

\subsubsection{Quasi-Monte Carlo - Sobol Sequences}\label{subsubsec:qmc}
The sample size has a significant impact on both the computational cost of each iteration and the quality of the model. Therefore, finding a balance between sample size and accuracy is imperative to large-scale applications. Contemporary methods for sample selection in metal additive manufacturing simulation utilize multiple uniformly sampled grids of varying resolution to induce bias~\citep{liPhysicsinformedNeuralNetwork2023, liaoHybridThermalModeling2023}. These methods, while effective in smaller domains, fail to scale well. With repetitive uniform-grid sampling, the number of required samples increases by $m^d$ as the domain increases by $m$. With this sampling method, increasing our test domain into a reasonable range for large-scale manufacturing would necessitate at least $10^3$ times more samples per time step. Sobol' sequences, alternatively, generate low-discrepancy samples that are uniformly distributed over the unit hypercube \citep{sobolDistributionPointsCube1967}. They are first-order collapsible, meaning that projections in lower dimensions retain uniform coverage. In the context of this work, Sobol' sequences ensure that spatial samples vary across time steps, which contrasts with uniform-grid implementations which repeat the same sampling pattern at every time step. Thus, more effectively covering the entire spatiotemporal domain, improving overall sampling uniformity.
\begin{figure}[htbp]
    \centering
    \begin{subfigure}[t]{0.49\linewidth}
    \includegraphics[trim={8cm 6cm 6cm 6.4cm}, clip, width=1\linewidth]{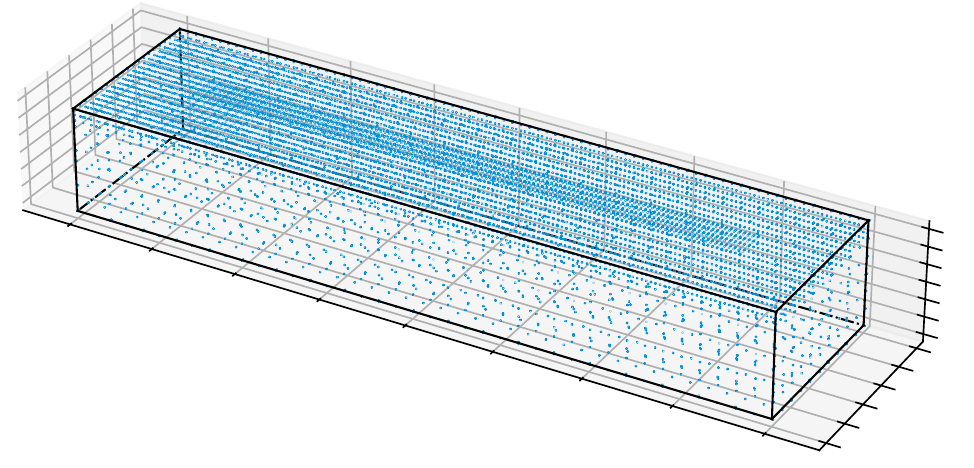}
    \caption{Uniform multi-grid sampling}
    \label{fig:uniform_sampling}
    \end{subfigure}
    \begin{subfigure}[t]{0.49\linewidth}
    \includegraphics[trim={8cm 6cm 6cm 6.4cm}, clip, width=1\linewidth]{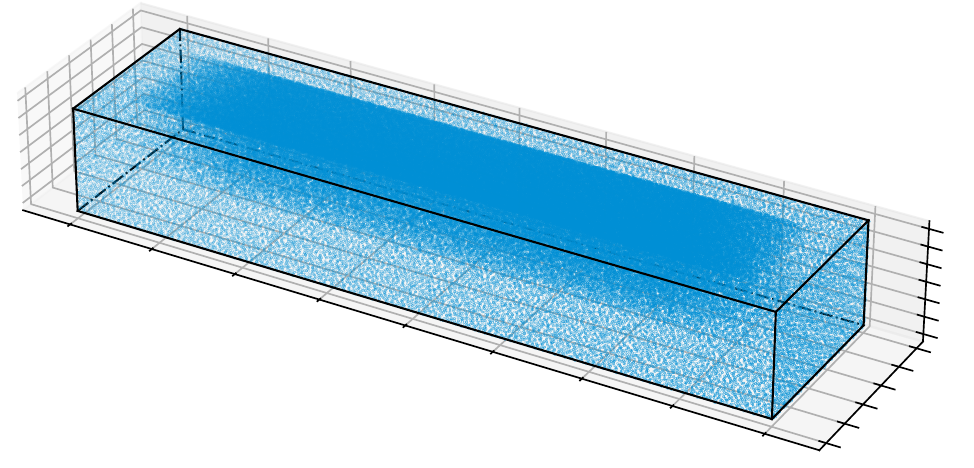}
    \caption{Sobol' sequence sampling}
    \label{fig:quasi_monte_carlo}
    \end{subfigure}
\caption{Example collocation point sampling comparison with collapsed time dimension}
\end{figure}
Despite appearing sparser, the number of samples in \Cref{fig:uniform_sampling} is more than double that of \Cref{fig:quasi_monte_carlo}. According to preliminary testing on current hardware, the expected computation time is linearly related to the sample size, resulting in approximately half of the necessary computation time per epoch while maintaining similar accuracy. The ratio between the necessary uniform samples and the Sobol' samples only increases as the spatial-temporal domain size increases, and therefore becomes increasingly more efficient in large-scale simulations. This is especially relevant for parameterized neural networks~\cite{choParameterizedPhysicsinformedNeural2024,LIU2024104937}, where dimensionality exceeds the four spatial-temporal inputs ($x,y,z,t$). For a given target integration error $\epsilon$, approximately $C_{MC}\cdot\epsilon^{-2}$ uniform samples are necessary, as compared with $C_{s}\cdot\epsilon^{-1} \bigl(\ln(\epsilon^{-1})\bigr)^{d}$ Sobol' samples \citep{niederreiterLowdiscrepancyLowdispersionSequences1988}. This, however, does not account for the quality of samples used in the training process, as by not overlapping samples, the number of unique samples is effectively multiplied by the number of time steps. Future refinement in this area could investigate other data-free sampling methods, such as Latin Hypercube sampling~\cite{mckay1979comparison}, Halton sequences~\cite{halton1960efficiency}, or adaptive sampling methods, such as Taylor-Expansion based Adaptive Design (TEAD)~\cite{moTaylorExpansionBasedAdaptive2017} or Monte Carlo Intersite-Projected Threshold (MIPT)~\cite{crombecqEfficientSpacefillingNoncollapsing2011}, among many others.

{\renewcommand{\arraystretch}{0.9}
\begin{table}[htbp]
    \caption{Network and Training Parameters}
    \vspace{-1.5mm}
    \centering
    \begin{tblr}{l l r}
        \toprule
                                  & \textit{Parameter}      & \textit{Value}               \\
        \midrule
        \SetCell[r=2]{m,2cm}{Topology} & Depth & 4\\
                                  & Width & 64\\
        \midrule
        Optimizer & & \\
        \midrule
         \SetCell[r=3]{m,2cm}{Adam}   & Learning Rate       &  0.001\\ 
                                 & Betas               & (0.9, 0.99)\\
                                      & Epochs         & 14850      \\
        \midrule[dotted]
        \SetCell[r=4]{m,2cm}{L-BFGS}  & Learning Rate       & 0.01        \\
                                 & Max. Iterations     & 50          \\ 
                                 & Max. Evaluations    & 62          \\ 
                                 & Epochs              & 150\\
        
        \bottomrule
    \end{tblr}
    \label{tab:network_training_parameters}
\end{table}
}

\subsubsection{Parameter Analysis and Tuning}
Substantial parameter analysis was performed to determine the optimal parameters. Additionally, various methodologies were implemented and tested, including, governing equation normalization~\cite{pengMultilayerThermalSimulation2024a}, random weight factorization (RWF)~\cite{wangRandomWeightFactorization2022}, random Fourier feature embeddings~\cite{tancikFourierFeaturesLet}, and a variety of learning rate schedulers; piecewise-constant training, stochastic gradient descent~\cite{loshchilovSGDRSTOCHASTICGRADIENT2017}, and one-cycle super-convergence~\cite{smithSuperConvergenceVeryFast2018}. However, these methods did not provide significant benefits in terms of accuracy or computational efficiency. The chosen parameters, detailed in \Cref{tab:network_training_parameters}, were the most effective and consistent methods for our circumstances.

\subsection{Comparison with FEM}\label{subsec:comparison}
\begin{figure}[htbp]
    \centering
    \begin{subfigure}[b]{1\linewidth}
        \centering
        \includegraphics[trim={0 11.75cm 0 0.5cm}, clip, width=1\linewidth]{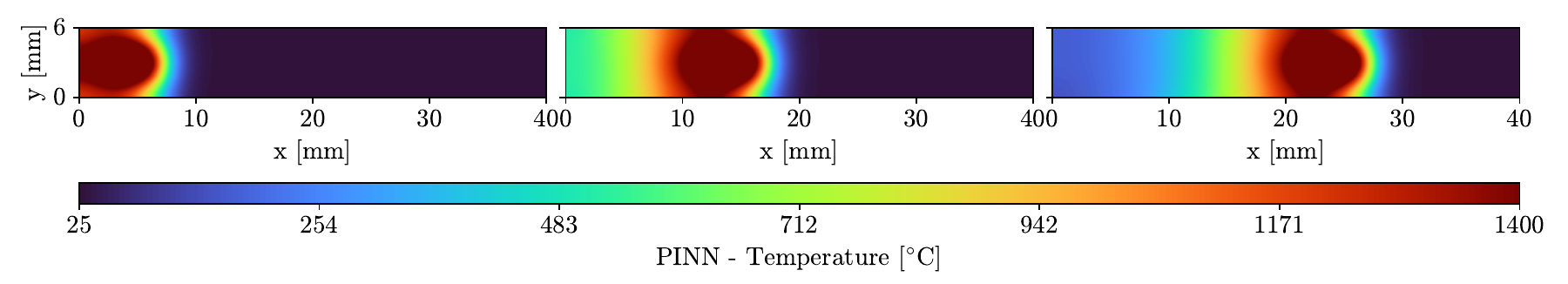}
        \caption{PINN model prediction output for t = 0.6~s, 1.6~s, 2.6~s}
    \end{subfigure}
    \hfill
    \begin{subfigure}[b]{1\linewidth}
        \centering
        \includegraphics[trim={0 6cm 0 6.5cm}, clip, width=1\linewidth]{figures/results/case1/fine/full_comparison_z=4.0.pdf}
        \caption{FEM-FC model simulation output for t = 0.6~s, 1.6~s, 2.6~s}
    \end{subfigure}
    \hfill
    \begin{subfigure}[b]{1\linewidth}
        \centering
        \includegraphics[trim={0cm 0.5cm 0 12cm}, clip, width=1\linewidth]{figures/results/case1/fine/full_comparison_z=4.0.pdf}
        \caption{Absolute relative difference between FEM-FC and PINN Temperatures}
    \end{subfigure}
    
    \caption{Comparison of the temperature field between FEM-FC and trained PINN model on the top surface (z=4~mm)}
    \label{fig:full_comparison_case_1_z}
\end{figure}

\begin{figure}[htbp]
    \centering
    \begin{subfigure}[b]{1\linewidth}
        \centering
        \includegraphics[trim={0 9.20cm 0 0}, clip, width=1\linewidth]{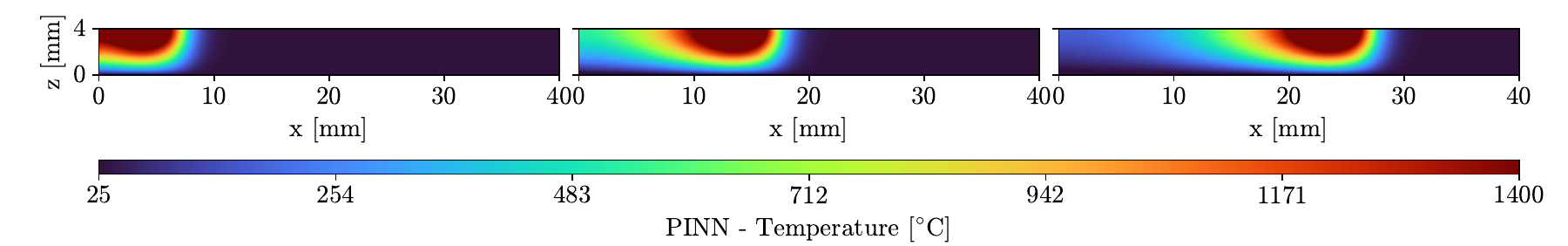}
        \caption{PINN model prediction output for t = 0.6~s, 1.6~s, 2.6~s}
    \end{subfigure}
    \hfill
    \begin{subfigure}[b]{1\linewidth}
        \centering
        \includegraphics[trim={0 4.75cm 0 4.75cm}, clip, width=1\linewidth]{figures/results/case1/fine/full_comparison_y=3.0.pdf}
        \caption{FEM-FC model simulation output for t = 0.6~s, 1.6~s, 2.6~s}
    \end{subfigure}
    \hfill
    \begin{subfigure}[b]{1\linewidth}
        \centering
        \includegraphics[trim={0 0.25cm 0 9.2cm}, clip, width=1\linewidth]{figures/results/case1/fine/full_comparison_y=3.0.pdf}
        \caption{Absolute relative difference between FEM-FC and PINN Temperatures}
    \end{subfigure}
    \caption{Comparison of the temperature field between FEM and trained PINN model on the y-midsection surface (y=3~mm)}
    \label{fig:full_comparison_case_1_y}
\end{figure}

In this section, we present a comprehensive comparison between the PINN thermal history predictions and the FEM model results, which are taken as our ground truth. To reiterate, the PINN model was trained entirely without the use of external data, solely using the normalized weighted total sum calculated using the residual errors based on the governing equations. For this comparison, we primarily examine two metrics, the relative $L^2$ error and the runtime. A trained PINN inherently exhibits super-resolution properties, which means it is not limited by the spatial or temporal discretization for which it was trained. Any point within the spatial-temporal domain can be predicted, regardless of whether it was included in the original set of collocation points. Thus, it is challenging to select the discretization for the FEM mesh that provides the best comparison with the PINN results, particularly in terms of computational cost. To evaluate the computational efficiency of the PINN, two finite element models were constructed, as detailed in \Cref{sec:fem}, and utilized in three simulations with progressively increasing spatial and temporal resolutions. These simulations are hereafter referred to as FEM-CC, FEM-FC, and FEM-FF. The FEM-CC model uses a uniform spatial mesh of 0.5~mm and a temporal resolution of 20~ms. The FEM-FC model adopts a finer spatial mesh of 0.1~mm while maintaining the same temporal resolution of 20~ms. The FEM-FF model further refines the temporal resolution to 5~ms, retaining the 0.1~mm spatial mesh. The FEM-FF model most closely resembles the discretization on which the PINN was trained. 
As shown in \Cref{fig:full_comparison_case_1_z,fig:full_comparison_case_1_y}, the FEM and PINN results are presented for three distinct time steps, 0.6~s, 1.6~s, and 2.6~s, corresponding to the spatial-temporal resolution of the FEM-FC simulation. At all time-steps, both models exhibit consistent alignment throughout, with the highest absolute percent relative error at the leading edge of the heat source, where temperature values are relatively low and thus minor discrepancies are amplified. The maximum temperature has been capped at 1400$^\circ$C to highlight the melt pool. The relative $L^2$ error, calculated across the entire domain, including inter-melt pool regions, was $7.267 \times 10^{-2}$. The computation times for the PINN training and the FEM simulation can be seen in \Cref{tab:computation_time}. Apart from the coarsest FEM model, the PINN training requires significantly less computational effort than the fine and super-fine meshes at a 62.4\%  and 98.6\% reduction, respectively.

\begin{table}[hbpt]
    \centering
    \caption{Computational Time Comparison}
    \vspace{-1.5mm}
    \begin{tblr}{c l r}
        \toprule
        \SetCell[c=2]{m}{\textit{Method}} &   & \textit{Time}\\
        \midrule
        \SetCell[c=2]{l}{PINN}   &            &         45m: 40.0s\\
        \midrule[dotted]
        \SetCell[r=3]{l}{FEM}    & CC         &         01m: 33.0s\\
                                 & FC         &     2h: 01m: 22.5s\\
                                 & FF         & 2d: 6h: 21m: 15.0s\\
         \bottomrule
    \end{tblr}
    \label{tab:computation_time}
\end{table}

\begin{figure}[hbpt]
    \centering
    \begin{subfigure}[b]{0.7\linewidth}
    \centering
    \includegraphics[trim={0 0 1cm 1cm}, clip, width=\linewidth]{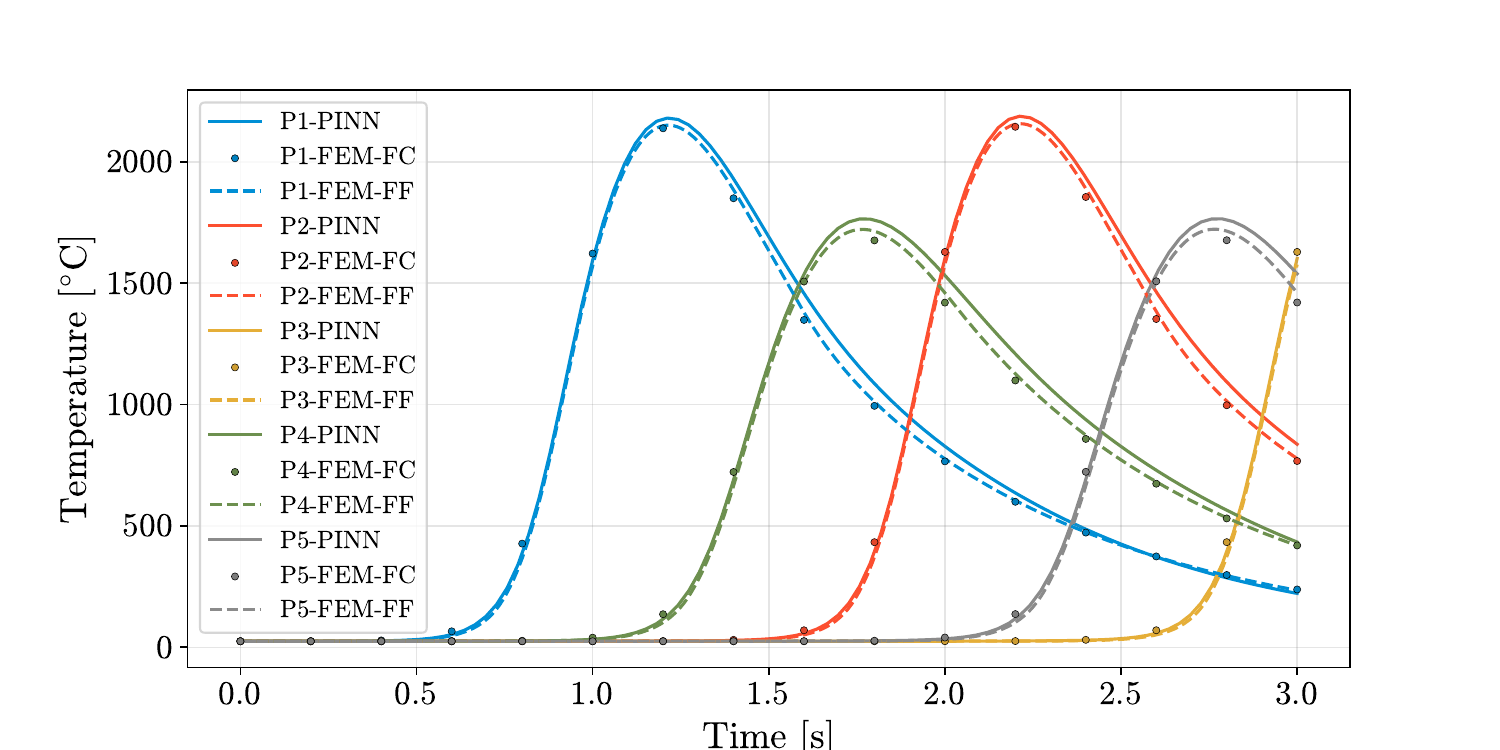}
    \caption{Thermal histories comparison between FEM-FC, FEM-FF, and PINN}
    \end{subfigure}
    \hfill
    \begin{subfigure}[b]{0.65\linewidth}
    \centering
    \includegraphics[width=0.95\linewidth]{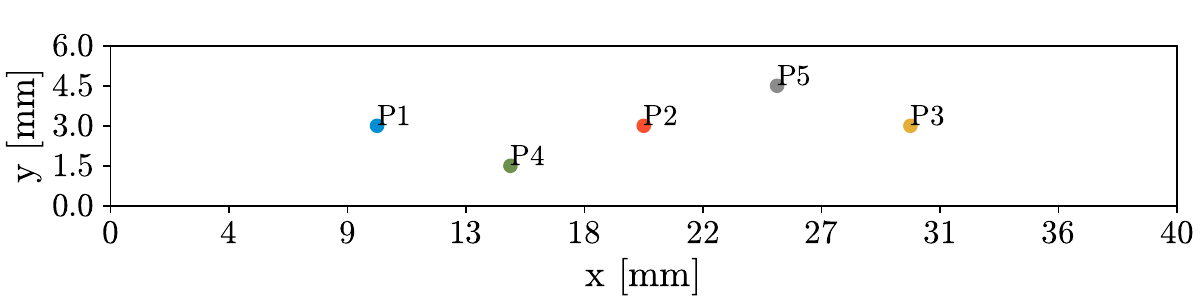}
    \caption{Representative point locations at top-surface (+z)}
    \end{subfigure}
    \caption{Thermal history comparison of representative points on the top-surface (z=4~mm)}
    \label{fig:temp_over_time_case_1}
\end{figure}

\begin{figure}[hbpt]
    \centering
    \begin{subfigure}[t]{0.49\linewidth}
        \includegraphics[trim={2.8cm 0.0cm 20.65cm 1cm}, clip,width=1.\linewidth]{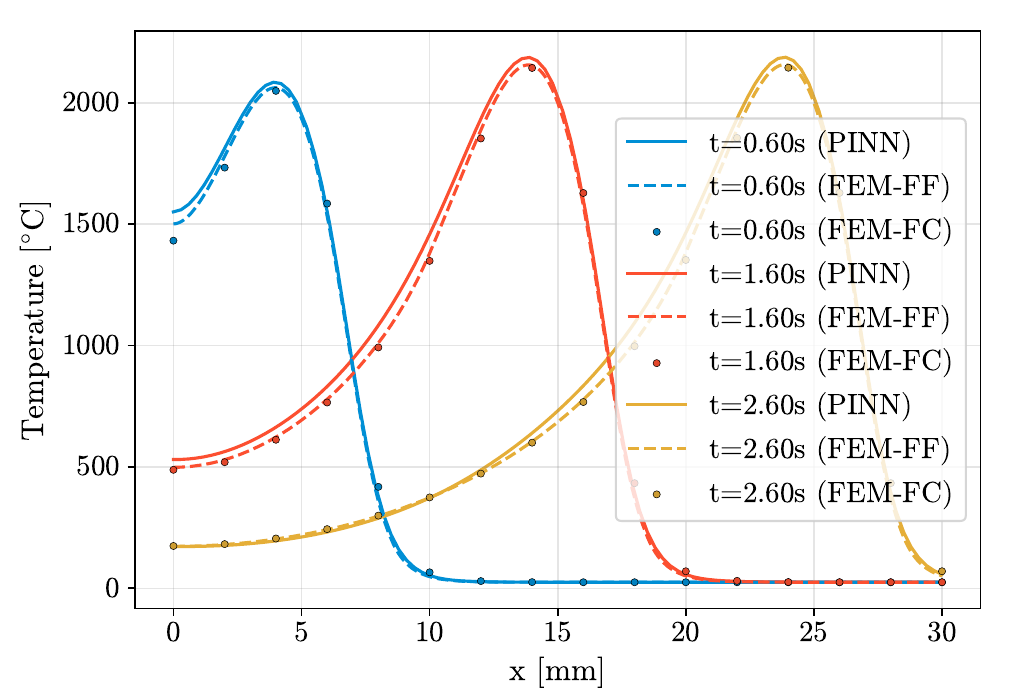}
        \caption{FEM and PINN temperature distribution along heat source path (x=3~mm) on the top-surface (The red dashed line in \Cref{fig:case1_schematic})}
    \end{subfigure}
    \hfill
    \begin{subfigure}[t]{0.49\linewidth}
        \includegraphics[trim={20cm 0cm 3.25cm 1cm}, clip,width=1.\linewidth]{figures/results/case1/fine/temp_along_tracks_smooth.pdf}
        \caption{Absolute temperature difference between FEM-FF and PINN}
    \end{subfigure}
    \caption{Thermal comparison between FEM-FC, FEM-FF, and PINN at t = 0.6~s 1.6~s and 2.6~s}
    \label{fig:temp_along_path_case_1}
\end{figure}

Five representative points were selected along the top surface, both on and off the path of the heat source. The results, seen in \Cref{fig:temp_over_time_case_1}, show minimal discrepancy between FEM and PINN results, particularly along the heat source path, again highlighting the limitations of FEM. Without interpolation, the verifiable thermal history is limited to discrete points in the spatial-temporal domain, and, as seen in \Cref{tab:computation_time}, increasing the fidelity can have computationally infeasible consequences. The accuracy along the entire heat source path was compared at multiple time-steps in \Cref{fig:temp_along_path_case_1}, which shows an average maximum temperature difference of approximately 85$^\circ$C or 6.5\%. Given that the FEM benchmark itself entails numerical errors, the PINN predictions can be regarded as being in good agreement and, consequently, validated.

\subsection{Potential and Limitations}\label{subsec:limitations}
Nevertheless, due to the absence of external training data, this strictly PDE-based approach has inherent accuracy limitations, regardless of the iteration count or optimization strategies employed. However, by limiting the method to a strictly PDE-driven approach, the workflow becomes entirely independent of the computational time and efficiency constraints associated with external data sources. Therefore, provided it is sufficiently accurate and more computationally efficient, as has been shown, the trained PINN model can be considered a justifiable substitution for an alternative simulation or analytical counterpart. Data-driven techniques can only justify themselves over the methodology for which they were trained when the model can be reused for multiple use cases or if the total sum of training time and necessary external data is less than that of the comparable model. In addition, the resulting PINN models are highly memory-efficient, typically requiring only a few kilobytes of storage. In contrast, extracting and storing results from FEM simulations using software like Abaqus can require several gigabytes and take several hours to complete. This difference in storage and extraction time further emphasizes the practicality of PINNs for rapid, lightweight deployment and inference.

\newpage
\section{Conclusion}\label{sec:conclusion}
This study highlights the potential of PINN to outperform conventional FEM in terms of computational efficiency for thermal simulation of large-scale components relevant to structural engineering. Realizing this potential requires a careful design of the neural network by leveraging advanced strategies to enhance its computational efficiency. A primary focus of this study was on improving the sampling strategy for collocation points, which are a key computational bottleneck during training, particularly as the problem size increases. PINN provides a unique advantage over FEM results with its inherent "super resolution" capability, offering high-resolution predictions in both space and time, whereas FE analysis requires extra computational effort to achieve comparable resolution by necessitating finer meshes. The following conclusions can be drawn from this study:

\begin{itemize}
    \item PINN significantly reduced computational effort compared to FEM simulations, achieving a 62.5\% reduction for the FEM-FC mesh and 98.6 \% for the FEM-FF mesh. 
    \item Being trained solely on PDE without external data, the PINN model achieved a relative $L^2$ error of $7.267 \times 10^{-2}$ compared to FEM benchmarks.
    \item Employing Sobol' sequences for sampling collocation points effectively reduced computational time while maintaining desired accuracy.
    
\end{itemize}

Although this study maintained a constant geometry, the exploration of various resolutions in FE analysis versus super-resolution of PINN elucidates the scalability potential of PINNs, as if the resolution were constant while the domain size increased. While auxiliary data is reported to facilitate training \cite{liaoHybridThermalModeling2023}, relying solely on PDE without any external data is inevitable to obtain improved computational efficiency over FEM; involving results of even one FEM simulation in training makes the total costs associated with PINN more than that of a single FEM simulation. This holds true when PINNs are developed for specific cases. However, a noteworthy prospect is the development of parametric PINNs, trained across a range of inputs. Such models could be reused multiple times with different sets of inputs at minimal computational cost after initial training, which can justify investing in even higher computational costs for training. A preliminary version of this work has been made available as a preprint on arXiv.

\section*{CRediT author contribution statement}
\textbf{Michael Ryan:} Writing - review \& editing, Writing - original draft, Visualization, Software, Validation, Methodology, Conceptualization. \textbf{Mohammad Hassan Baqershahi:} Writing - review \& editing, Writing - original draft, Validation, Methodology, Conceptualization. \textbf{Hessamoddin Moshayedi:} Writing - review \& editing, Writing - original draft, Supervision, Validation, Methodology, Conceptualization. \textbf{Elyas Ghafoori:} Writing - review \& editing, Supervision, Validation, Methodology, Conceptualization, Funding acquisition.

\section*{Declaration of competing interest}
The authors declare that they have no known competing financial interests or personal relationships that could have appeared to influence the work reported in this paper.

\section*{Acknowledgment}
This work was supported by funding from Niedersächsisches Ministerium für Wissenschaft und Kultur (MWK), Germany, with the grant number ZN3725.

\bibliographystyle{elsarticle-num} 
\bibliography{references.bib}

\end{document}